\documentclass[useAMS,usenatbib]{mn2e}
\usepackage{mn2e-breakabs}
\usepackage{graphicx}
\usepackage{subfigure}
\usepackage{times}
\usepackage{multicol}
\newcommand{\Hunit}{\,{\rm km}\,{\rm s}^{-1}\,{\rm Mpc}^{-1}}

\def\la{\mathrel{\mathpalette\fun <}}
\def\ga{\mathrel{\mathpalette\fun >}}
\def\fun#1#2{\lower3.6pt\vbox{\baselineskip0pt\lineskip.9pt
        \ialign{$\mathsurround=0pt#1\hfill##\hfil$\crcr#2\crcr\sim\crcr}}}

\def\bfs{\mbox{\bf s}}

\newcommand{\be}{\begin{equation}}
\newcommand{\ee}{\end{equation}}
\newcommand{\ba}{\begin{eqnarray}}
\newcommand{\ea}{\end{eqnarray}}
\newcommand{\simgt}{\,\hbox{\lower0.6ex\hbox{$\sim$}\llap{\raise0.6ex\hbox{$>$}}}\,}
\newcommand{\simlt}{\,\hbox{\lower0.6ex\hbox{$\sim$}\llap{\raise0.6ex\hbox{$<$}}}\,}

\begin{document}

\title[$H(z)$, $D_A(z)$, and $f(z)\sigma_8(z)$ from SDSS DR7 LRGs]
{Modeling the Anisotropic Two-Point Galaxy Correlation Function on Small Scales
and Single-Probe Measurements of $H(z)$, $D_A(z)$, and $f(z)\sigma_8(z)$ from the 
Sloan Digital Sky Survey DR7 Luminous Red Galaxies}

\author[Chuang \& Wang]{
  \parbox{\textwidth}{
 Chia-Hsun Chuang$^1$\thanks{MultiDark Fellow; E-mail: chia-hsun.chuang@uam.es}
 and Yun Wang$^2$
  }
  \vspace*{4pt} \\
  $^1$ Instituto de F\'{\i}sica Te\'orica, (UAM/CSIC), Universidad Aut\'onoma de Madrid,  Cantoblanco, E-28049 Madrid, Spain \\
  $^2$ Homer L. Dodge Department of Physics \& Astronomy, Univ. of Oklahoma,
                 440 W Brooks St., Norman, OK 73019, U.S.A.\\
}

\date{\today} 

\maketitle

\begin{abstract}

We present a simple and efficient phenomenological model for the two-dimensional two-point 
galaxy correlation function that works well over a wide range of scales, from large scales down to 
scales as small as 25$\,h^{-1}$Mpc. Our model incorporates nonlinear effects, a scale-dependent 
galaxy bias on small scales, and allows the redshift-space distortions to be scale and direction 
dependent. We validate our model using LasDamas mock catalogs, and apply it to the 
Sloan Digital Sky Survey (SDSS) DR7 Luminous Red Galaxies (LRGs).
Using only the monopole and quadrupole of the correlation function measured from the SDSS DR7 LRGs,
we obtain improved measurements $H(z)r_s(z_d)/c=0.0433\pm 0.0042$, $D_A(z)/r_s(z_d)=6.59\pm 0.46$, and 
$f(z)\sigma_8(z)=0.429\pm 0.089$ at $z=0.35$, using the scale range of $25<s<120h^{-1}$Mpc.
We expect our results and model to be useful in tightening dark energy and gravity constraints
from the full analysis of current and future galaxy clustering data.

\end{abstract}

\begin{keywords}
  cosmology: observations, distance scale, large-scale structure of
  Universe
\end{keywords}

\section{Introduction}

In our quest to solve the mystery of the observed cosmic acceleration
\citep{Riess:1998cb,Perlmutter:1998np}, galaxy clustering plays an increasingly 
important role as a probe of both dark energy and gravity, the two
main classes of possible explanations for cosmic acceleration.
Current data from the Sloan Digital Sky Survey (SDSS) Data Release Seven (DR7) 
\citep{Abazajian:2008wr}, WiggleZ \citep{Blake09}, and BOSS \citep{Eisenstein11}
are allowing us to place very useful constraints on dark energy.
The planned space mission Euclid\footnote{http://www.euclid-emc.org/}
will survey $\sim$ 60 million emission-line galaxies at $0.7<z<2$ over 15,000 square degrees 
\citep{Cimatti:2008kc,Wang10,RB}, and provide potentially revolutionary 
bounds on the nature of cosmic acceleration.

The SDSS data have been analyzed using both the power spectrum 
method (see, e.g., \citealt{Tegmark:2003uf,Hutsi:2005qv,Padmanabhan:2006ku,Blake:2006kv,Percival:2007yw,Percival:2009xn,Reid:2009xm,Montesano:2011bp}), 
and the correlation function method (see, e.g., 
\citealt{Eisenstein:2005su,Okumura:2007br,Cabre:2008sz,Martinez:2008iu,Sanchez:2009jq,Kazin:2009cj,Chuang:2010dv,Samushia:2011cs,Padmanabhan:2012hf}). 
Although these two methods are simple Fourier transforms of one another, the analysis processes are quite 
different and the results cannot be converted using Fourier transform 
directly because of the finite size of the survey volume. 

The power of galaxy clustering as a dark energy probe
lies in the fact that the Hubble parameter, $H(z)$, the
angular diameter distance, $D_A(z)$, can in principle
be extracted simultaneously from data through the measurement
of the baryon acoustic oscillation (BAO) scale 
in the radial and transverse directions \citep{BG03,SE03,Wang06}. 
The inclusion of information from full galaxy clustering
goes beyond BAO only, and enables significantly enhanced constraints on
$H(z)$ and $D_A(z)$. Most importantly, it allows the measurement of
the growth rate of cosmic large scale structure, $f(z)=\beta(z) b(z)$ (where
$\beta(z)$ denotes the linear redshift-space distortion (RSD) factor \citep{Kaiser:1987qv}, 
and $b(z)$ denotes galaxy bias), required for using
galaxy clustering to test gravity \citep{Guzzo08,Wang08}.

In fact, it is possible to measure $f(z)\sigma_8(z)$ \citep{Song09}
or $f(z) \sigma_m(z)/r_s(z_d)^4$ \citep{Wang12} without facing the difficulty of measuring galaxy bias.

In \cite{Chuang:2011fy}, we made significant improvements
in modeling galaxy clustering from previous studies \citep{Okumura:2007br,Cabre:2008sz,Kazin:2010nd},
and succeeded in making the first simultaneous measurements of $H(z)$ and $D_A(z)$ from data, using
the full 2D correlation function of a sample of SDSS DR7 LRGs \citep{Eisenstein:2001cq}, and 
without assuming a dark energy model or a flat Universe. 
\cite{Xu:2012fw} measured $H(z)$ and $D_A(z)$ at $z=0.35$ from the
SDSS DR7 LRGs by applying density-field reconstruction to an anisotropic analysis of the BAO peak.
\cite{Anderson:2013oza} applied the same method on SDSS III Baryon Oscillation 
Spectroscopic Survey (SDSS-III/BOSS) DR9 sample.
Regarding the measurements of growth constraints, \cite{Samushia:2011cs} measured $f(z)\sigma_8(z)$ from 
SDSS DR7 LRG sample with CMB + SNIa priors.
\cite{Reid12} measured $H(z)$, $D_A(z)$, and $f(z)\sigma_8(z)$
at $z=0.57$ from the monopole and quadrupole of the 2D 2PCF of the SDSS-III/BOSS DR9 sample, assuming CMB priors.
Most recently, \cite{Chuang:2013hya} applied similar analysis as this paper on SDSS-III/BOSS DR9 sample to
measure $H(z)$, $D_A(z)$, $\Omega_m h^2$ and $f(z)\sigma_8(z)$ without CMB priors.

In \cite{Chuang:2012ad}, we extended our method by exploring the use of the multipoles of the 
correlation function to measure $H(z)$, $D_A(z)$, and $\beta(z)$.
The obvious advantage of using multipoles of the correlation function instead of
the full 2D correlation function is the reduced number of data points used to obtain 
similar amount of information.

The proper modeling of RSD is required in order to measure $\beta(z)$ or $f(z)$
from galaxy clustering data. Recent work on improving the modeling of RSD
include that of \cite{Jennings:2010uv} and \cite{Reid:2011ar}.
In this paper, we focus on the detailed phenomenological modeling of the correlation function
on smaller scales to obtain improved constraints on $\beta(z)$ or $f(z)\sigma_8(z)$.
We use the multipoles of the 2D correlation function for speed and efficiency.

In Section \ref{sec:data}, we introduce the galaxy sample used 
in our study. In Section \ref{sec:model}, we describe the details of our 
new model. In Section \ref{sec:method}, we describe the details of our methodology.
In Section \ref{sec:results_sdss}, we present our improved measurements from SDSS DR7 LRGs.
We summarize and conclude in Sec.~\ref{sec:conclusion}.

\section{Data} \label{sec:data}

The SDSS has observed one-quarter of the
entire sky and performed a redshift survey of galaxies, quasars and
stars in five passbands $u,g,r,i,$ and $z$ with a 2.5m telescope
\citep{Fukugita:1996qt,Gunn:1998vh,Gunn:2006tw}. 
We use the public catalog, the NYU Value-Added Galaxy Catalog
(VAGC) \citep{Blanton:2004aa}, derived from the
SDSS II final public data release, Data Release 7 (DR7)
\citep{Abazajian:2008wr}.
We select our LRG sample from the NYU VAGC with 
the flag $primTarget$ bit mask set to $32$. K-corrections
have been applied to the galaxies with a
fiducial model ($\Lambda$CDM with $\Omega_m=0.3$ and $h=1$), and
the selected galaxies are required to have rest-frame $g$-band absolute
magnitudes $-23.2<M_g<-21.2$ \citep{Blanton:2006kt}. The same 
selection criteria were used in previous papers
\citep{Zehavi:2004zn,Eisenstein:2005su,Okumura:2007br,Kazin:2009cj}. 
The sample we use is referred to as ``DR7full'' in \cite{Kazin:2009cj}.
Our sample includes 87000 LRGs in the redshift range 0.16-0.44.

Spectra cannot be obtained for objects closer than 55 arcsec
within a single spectroscopic tile due to the finite size of the
fibers. To correct for these ``collisions'', the redshift of an object
that failed to be measured would be assigned to be the same as the
nearest successfully observed one. Both fiber 
collision corrections and K-corrections have been made in NYU-VAGC 
\citep{Blanton:2004aa}. The collision corrections applied here are 
different from what has been suggested in \cite{Zehavi:2004zn}. 
However, the effect should be small since we are using relatively large 
scale which are less affected by the collision corrections.

We construct the radial selection function as a cubic spline fit 
to the observed number density histogram with the width $\Delta
z=0.01$. The NYU-VAGC provides the description of the
geometry and completeness of the survey in terms of spherical
polygons. We adopt it as the angular selection function of our
sample. We drop the regions with completeness
below $60\%$ to avoid unobserved plates \citep{Zehavi:2004zn}. The 
Southern Galactic Cap region is also dropped.

\section{Modeling 2D Correlation Function} 
\label{sec:model}

In this section, we describe our model which encompasses the linear scale to the nonlinear scale.

\subsection{Modeling 2D Correlation Function for Large Scales}
\label{sec:model_large}

We compute the linear matter power spectra, $P_{lin}(k)$, by using CAMB \citep{Lewis:1999bs}. The linear power spectrum can be composed to two parts:
\begin{equation} \label{eq:pk_lin}
P_{lin}(k)=P_{nw}(k)+P_{BAO}^{lin}(k),
\end{equation}
where $P_{nw}(k)$ is the no-wiggle or pure CDM power spectrum calculated using Eq.(29) from \cite{Eisenstein:1997ik} and $P_{BAO}^{lin}(k)$ is the wiggled part defined by the equation itself.
The nonlinear damping effect of the wiggled part in redshift space can be well approximated by \citep{Eisenstein:2006nj}
\begin{equation} \label{eq:bao}
P_{BAO}^{nl}(k,\mu_k)=P_{BAO}^{lin}(k)\cdot \exp\left(-\frac{k^2}{2k_\star^2}[1+\mu_k^2(2f+f^2)]\right),
\end{equation}
where $k_\star$ could be computed by \citep{Crocce:2005xz, Matsubara:2007wj}
\begin{equation} \label{eq:kstar}
k_\star=\left[\frac{1}{3\pi^2}\int P_{lin}(k)dk\right]^{-1/2}.
\end{equation}
The dewiggled power spectrum is
\begin{equation} \label{eq:pk_dw}
P_{dw}(k,\mu_k)=P_{nw}(k)+P_{BAO}^{nl}(k,\mu_k),
\end{equation}
$\mu_k$ is the cosine of the angle between ${\bf k}$ and the line of sight (LOS).
Note that Eqs.(\ref{eq:pk_lin})-(\ref{eq:pk_dw}) are the same as Eq.(2) in
\cite{Chuang:2011fy}, except for the addition of the direction-dependent
terms in the exponent of the damping factor in Eq.(\ref{eq:bao}), but
are somewhat more intuitive.

Next, we include the linear RSD as follows to
obtain the galaxy power spectrum in redshift space at large scales \citep{Kaiser:1987qv}:
\begin{eqnarray} \label{eq:pk_2d}
P_g^s(k,\mu_k)&=&b^2(1+\beta\mu_k^2)^2P_{dw}(k,\mu_k),\\
              &=&P_{g,nw}^s(k,\mu_k)+P_{g,BAO}^s(k,\mu_k),\label{eq:P(k)_g_s}
\end{eqnarray}
where $b$ is the linear galaxy bias.
Note that we have defined
\ba
&& \hspace{-0.2in} P_{g,nw}^s(k,\mu_k) = b^2(1+\beta\mu_k^2)^2 P_{nw}(k) \label{eq:P_nw_s} \\
&& \hspace{-0.2in} P_{g,BAO}^s(k,\mu_k)= b^2(1+\beta\mu_k^2)^2 P_{BAO}^{nl}(k,\mu_k) \nonumber\\
=&& \hspace{-0.2in} b^2(1+\beta\mu_k^2)^2 P_{BAO}^{lin}(k)\cdot \exp\left(-\frac{k^2}{2k_\star^2}[1+\mu_k^2(2f+f^2]\right). \label{eq:P_bao_s}
\ea

%The model introduced above could be computed in the configuration space by Legendre polynomials expansions and integral convolutions.
Analogous to Eq.(\ref{eq:P(k)_g_s}), the galaxy correlation function can be decomposed 
into no-wiggle and wiggled parts as follows:
\begin{equation} \label{eq:xi_2d}
\xi_{g,dw}^s(\sigma,\pi)=\xi_{g,nw}^s(\sigma,\pi)+\xi_{g,BAO}^s(\sigma,\pi).
\end{equation}
While $\xi_{g,dw}^s(\sigma,\pi)$ can be obtained by Fourier-transforming
$P_g^s(k,\mu_k)$, doing so involves two-dimensional integrals, and thus is
time-consuming and inefficient. Instead, we can Fourier transform each term
in Eq.(\ref{eq:P(k)_g_s}) separately, using Legendre polynomial expansions and integral convolutions
that only involve one-dimensional integrals.

The no-wiggle galaxy correlation function in redshift space can be computed by 
Fourier transforming Eq.(\ref{eq:P_nw_s}), which gives \citep{hamilton1992}
\begin{equation} \label{eq:xi_nw}
\xi_{g,nw}^s(\sigma,\pi)=b^2(\xi_0^{nw}(s)P_0(\mu)+\xi_2^{nw}(s)P_2(\mu)+\xi_4^{nw}(s)P_4(\mu)),
\end{equation}
where $s=\sqrt{\sigma^2+\pi^2}$, 
$\mu$ is the cosine of the angle between $\bfs=(\sigma,\pi)$ and the LOS, and 
$P_l$ are Legendre polynomials. The multipoles of $\xi^{nw}$ are defined as
\begin{eqnarray} \label{eq:xi_mp}
 \xi_0^{nw}(r)&=&\left(1+\frac{2\beta}{3}+\frac{\beta^2}{5}\right)\xi^{nw}(r),\\
 \xi_2^{nw}(r)&=&\left(\frac{4\beta}{3}+\frac{4\beta^2}{7}\right)[\xi^{nw}(r)-\bar{\xi}(r)],\\
 \xi_4^{nw}(r)&=&\frac{8\beta^2}{35}\left[\xi^{nw}(r)+\frac{5}{2}\bar{\xi}^{nw}(r)
 -\frac{7}{2}\overline{\overline{\xi}}^{nw}(r)\right],
\end{eqnarray}
where $\beta$ is the linear RSD parameter and
\begin{eqnarray} \label{eq:xi_bar}
 \bar{\xi}^{nw}(r)&=&\frac{3}{r^3}\int_0^r\xi^{nw}(r')r'^2dr',\\
 \overline{\overline{\xi}}^{nw}(r)&=&\frac{5}{r^5}\int_0^r\xi^{nw}(r')r'^4dr', \label{eq:xi_barbar}
\end{eqnarray}
where $\xi^{nw}(r)$ is obtained by fourier transforming $P_{nw}(k)$.

The wiggled part of the galaxy correlation function in redshift space, 
$\xi_{g,BAO}^s(\sigma,\pi)$, is obtained by Fourier transforming Eq.(\ref{eq:P_bao_s}).
Note that the $\mu$-dependent damping factor in $k$-sapce in Eq.(\ref{eq:P_bao_s}) 
becomes a Gaussian convolution in configuration space:
\begin{equation} \label{eq:xi_gBAO}
 \xi_{g,BAO}^s(\sigma,\pi)=\int_{-\infty}^\infty \xi^\star(\sigma,\pi-x)\, f_{\star}(x)dx,
\end{equation}
where $\xi^{\star}(\sigma,\pi)$ is the Fourier transform of 
$b^2(1+\beta\mu_k^2)^2 P_{BAO}^{lin}(k)\cdot \exp(-\frac{k^2}{2k_\star^2})$, and 
\begin{equation}
 f_\star(x)=\frac{1}{\sigma_\star\sqrt{\pi}}\exp\left(-\frac{x^2}{\sigma_\star^2}\right),
\end{equation}
where 
\begin{equation}
\sigma_\star^2=\frac{4f+2f^2}{k_\star^2}.
\end{equation}
$\xi^{\star}(\sigma,\pi)$ can be obtained using Eq. (\ref{eq:xi_nw})-(\ref{eq:xi_barbar}),
but replace $\xi^{nw}(r)$ (the Fourier transform of $P_{nw}(k)$)
with the Fourier transform of $P_{BAO}^{lin}(k)\cdot \exp(\frac{-k^2}{2k_\star^2})$. 

\begin{figure}
\centering
\includegraphics[width=0.8 \columnwidth,clip,angle=-90]{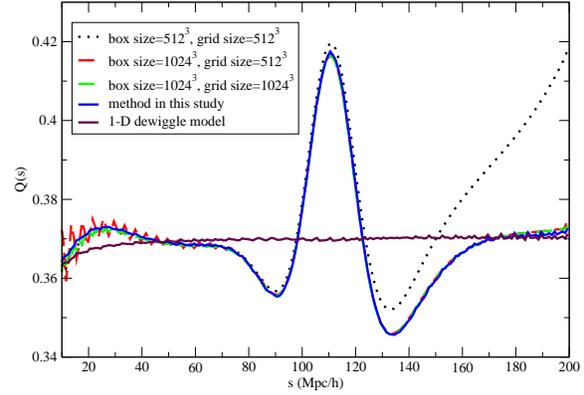}
\caption{The normalized quadrupoles from the correlation functions computed with FFT and our convolution method.
One can see that the results from FFT are converging to the result of the convolution method. In addition,
FFT method just reaches reasonable convergence with box size = $1024^3$ (Mpc/h)$^3$ and grid size = $1024^3$ for the scales considered in this study.
One would need to increase the box size or grid size if one want to include other scales. 
We also plot the $Q(s)$ from one-dimensional dewiggle model for comparison. It is a constant since the only redshift distortion effect comes from the Kaiser boost.}
\label{fig:q}
\end{figure}

Table. \ref{table:fftw} shows the performance of our convolution method by comparing with
the results using fast Fourier transform (FFT) directly. One can see our method is much more efficient. 
The two-dimensional dewiggle model has a obvious feature at the BAO scale for the normalized quadrupole, $Q(s)$ (see \citealt{Samushia:2011cs}, where
\begin{equation}
 Q(s)=\frac{\xi_2(s)}{\xi_0(s)-(3/s^3)\int_0^s\xi_0(s')s'^2ds'}.
\label{eq:Q}
\end{equation}
Fig. \ref{fig:q} shows that the results from the FFT method converges to the one from the convolution method. 
While these tests are performed on a single machine, the grid size used for FFT method is limited by the memory size.
One can still see some fluctuations at small scales for the maximum grid size (=$1024^3$). Therefore, our method not only provide
a much faster way but also use much less resources to compute the theoretical model. 
With a multi-cores machine (FFT can only use single core unless the machine's memory is much larger), our method could be hundreds times faster than FFT method. 
It is crucial while doing Markov Chain Monte Carlo (MCMC) analysis.
	
\begin{table}
\begin{center}
\begin{tabular}{rrr}\hline
box size (Mpc/h)$^3$ &grid size &computing time (sec) \\ \hline
$512^3$&\ \ $512^3$&\ \  $41$ \\
$1024^3$&\ \ $512^3$&\ \  $41$ \\
$1024^3$&\ \ $1024^3$&\ \  $352$ \\
\hline
\multicolumn{2}{r}{method used in this study}&\ \ $5$ \\
\hline
\end{tabular}
\end{center}
\caption{
The time needed for computing the two-dimensional dewiggle model described in Sec. \ref{sec:model_large}. 
We test different methods: one is performing fast Fourier transform (FFT) using FFTW\protect\footnotemark library; 
the other is the convolution method described in Sec. \ref{sec:model_large} which is the method used in this study.
One can see that our method is much faster than FFT method.
} \label{table:fftw}
\end{table}	

\footnotetext{http://www.fftw.org/}

\subsection{Modeling 2D Correlation Function on Small Scales}
For small scales, we need to model three effects: the nonlinear matter correlation function, the scale-dependent galaxy bias, 
and the RSD from the random galaxy pairwise velocities. It is well known that the small scale galaxy correlation function is well 
described by a powerlaw \citep{peebles1980}. Since the galaxy correlation function is given by
$\xi^{nw}(r)$ on small scales, we model the combination of nonlinear matter correlation function and the 
scale-dependent galaxy bias at small scales by multiplying $\xi^{nw}(r)$ with the following factor
\begin{equation}
b_{nl}(r)=r^{b_A F(r)},
\end{equation}
where $b_A$ is a constant. $F(r)$ is a function which is close to 1 for small $r$ and close to 0 when $r$ is large;
we choose
\begin{equation}
F(r)=\frac{1}{1+\left(\frac{r}{b_B}\right)^{b_C}},
\end{equation}
where we choose $b_B=30h^{-1}$Mpc and $b_C=4$; these are motivated by the fact that the galaxy correlation function is a powerlaw  
at small scales (i.e. $s<15h^{-1}$Mpc) and the scale-dependent effects (including nonlinear effects and 
scale-dependent galaxy bias) are negligible at larger scales, $s>40h^{-1}$Mpc.
The overall scale-dependent effects are included when computing the no-wiggle galaxy correlation function by replacing 
$\xi^{nw}(r)$ with $\xi^{nw}(r)\times b_{nl}(r)$ in applying Eq. (\ref{eq:xi_nw})-(\ref{eq:xi_barbar}). 
The resultant correlation function is denoted as $\xi_{g,nw}^{s,nl}(\sigma,\pi)$.

We now obtain the 2D correlation function that incorporate nonlinear effects, galaxy bias, and linear RSD:
\begin{equation} \label{eq:xi_star}
\tilde{\xi}(\sigma,\pi)=\xi_{g,nw}^{s,nl}(\sigma,\pi)+\xi_{g,BAO}^s(\sigma,\pi),
\end{equation} 
where $\xi_{g,BAO}^s(\sigma,\pi)$ is given by Eq.(\ref{eq:xi_gBAO}).

Next, we convolve the 2D correlation function with the distribution function of 
random pairwise velocities, $f(v)$, to obtain the final model $\xi(\sigma,\pi)$ 
\citep{peebles1980}
\begin{equation} \label{eq:theory}
 \xi(\sigma,\pi)=\int_{-\infty}^\infty \tilde{\xi}\left(\sigma,\pi-\frac{v}{H(z)a(z)}
 \right)\,f(v)dv,
\end{equation}
where the random motions are represented by an exponential form 
\citep{ratcliffe1998,Landy:2002xg}
\begin{equation}
 f(v)=\frac{1}{\sigma_v(s',\mu'^2)\sqrt{2}}\exp\left(-\frac{\sqrt{2}|v|}{\sigma_v(s',\mu'^2)}\right),
\end{equation}
where $\sigma_v$ is the pairwise peculiar velocity dispersion, $s'^2=\sigma^2+(\pi-\frac{v}{H(z)a(z)})^2$ and $\mu'=\frac{1}{s'}(\pi-\frac{v}{H(z)a(z)})$. We find that the 2D correlation functions measured from LasDamas mocks can be well fitted by
\begin{equation}
 \sigma_v(s',\mu'^2)=\sigma_{v,0}(1+C_\mu\mu'^2)(1+c_{\sigma1} e^{-c_{\sigma2}\sigma^2}),
\end{equation}
where $\sigma_{v,0}$ is the dispersion corresponding to the truly random motion and $c_\mu$, $c_{\sigma1}$, and $c_{\sigma2}$ (with unit of Mpc$^{-2}h^2$) terms describe the dependence on direction
and separation. The $\sigma$-dependence is similar to that found by \cite{Cabre:2008sz}. They found that the 2D correlation functions 
from the MICE N-body simulations are fitted well with a pairwise velocity distribution which is large when $\sigma<5h^{-1}$Mpc. 
We have added the direction-dependent term, $c_\mu\mu'^2$, to model the high amplitude of $\hat{\xi}_4$ at small scales (see Fig. \ref{fig:four}).

\section{Methodology}\label{sec:method}
In this section, we present the methodology and results of testing our model described in the previous section.

\subsection{Mock Catalogs Used}
We use the 160 mock catalogs from the LasDamas 
simulations\footnote{http://lss.phy.vanderbilt.edu/lasdamas/} 
(McBride et al., in preparation) to test our model.
LasDamas provides mock catalogs matching SDSS main galaxy and LRG samples.
We use the LRG mock catalogs from the LasDamas gamma release with the same cuts as
the SDSS LRG DR7full sample, $-23.2<M_g<-21.2$ and $0.16<z<0.44$.
We have diluted the mock catalogs to match the radial selection function 
of the observational data by randomly selecting the mock galaxies according to the 
number density of the data sample. We calculate the multipoles of the correlation functions 
of the mock catalogs and construct the covariance matrix (see \cite{Chuang:2012ad} for details).

\subsection{Measuring the Two-Dimensional Two-Point Correlation Function}

We convert the measured redshifts of galaxies to comoving distances 
by assuming a fiducial model, $\Lambda$CDM with $\Omega_m=0.25$. 
We use the two-point correlation function estimator given by 
\cite{Landy:1993yu}:
\begin{equation}
\label{eq:xi_Landy}
\xi(\sigma,\pi) = \frac{DD(\sigma,\pi)-2DR(\sigma,\pi)+RR(\sigma,\pi)}{RR(\sigma,\pi)},
\end{equation}
where $\pi$ is the separation along the line of sight (LOS), $\sigma$ 
is the separation in the plane of the sky, DD, DR, and RR represent the normalized 
data-data, data-random, and random-random pair counts respectively in a given
distance range. The LOS is defined as the direction from the observer to the 
center of a pair. The bin size we use here is
$1 \, h^{-1}$Mpc$\times 1 \,h^{-1}$Mpc. 
The Landy and Szalay estimator has minimal variance for a Poisson
process. Random data are generated with the same radial
and angular selection functions as the real data. One can reduce the shot noise due
to random data by increasing the number of random data. The number
of random data we use is 10 times that of the real data. While
calculating the pair counts, we assign to each data point a radial
weight of $1/[1+n(z)\cdot P_w]$, where $n(z)$ is the radial
selection function and $P_w = 4\cdot 10^4$ $h^{-3}$Mpc$^3$ 
\citep{Eisenstein:2005su}.

\begin{figure}
\centering
\includegraphics[width=1 \columnwidth,clip,angle=0]{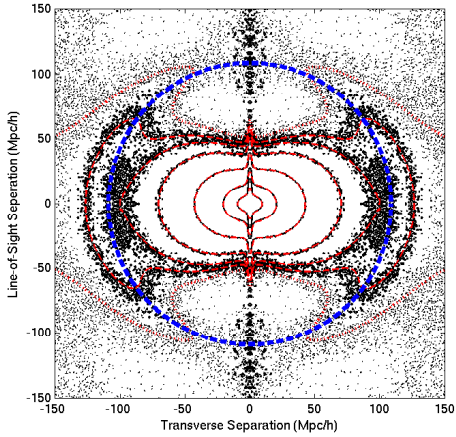}
\caption{The average two-dimensional two-point correlation
function (2D 2PCF) measured from 160 LasDamas SDSS LRGfull mock catalogs (solid black
contours), compared to a theoretical model with the input parameters 
of the LasDamas simulations and $\{\beta$, $b_A$, $\sigma_{v,0}$, $c_\mu$, $c_\sigma\}$ are set to $\{0.316$, $-0.0385$, $50$km\ s$^{-1}$, $10$, $4\}$ (dashed red contours).
The thick dashed blue circle denotes the baryon acoustic oscillation scale.
The contour levels are $\xi=2, 0.5, 0.1, 0.025, 0.01, 0.005, 0$.
The $\xi=0$ contours are denoted with dotted lines for clarity.}
\label{fig:twod}
\end{figure}

Fig \ref{fig:twod} shows the averaged 2D correlation function measured from the mock catalogs. 
We use the averaged radial selection function to construct the random catalog since it is closer to the true mean density.
Clearly, our model provides an excellent fit to data over a wide range of scales, from the largest scales where
data are not too noisy, to the smallest scales plotted (except very near the LOS).

\subsection{Multiples of the Correlation Function}
As in \cite{Chuang:2012ad}, the effective multipoles of the correlation function are defined by
\begin{equation}\label{eq:multipole}
 \hat{\xi}_l(s) \equiv \frac{\displaystyle\sum_{s-\frac{\Delta s}{2} < \sqrt{\sigma^2+\pi^2} < s+\frac{\Delta s}{2}}(2l+1)\xi(\sigma,\pi)P_l(\mu)\sqrt{1-\mu^2}}{\mbox{Number of bins used in the numerator}},
\end{equation}
where $\Delta s=5$ $h^{-1}$Mpc in this work, and 
\begin{equation}
\sigma=(n+\frac{1}{2})\mbox{$h^{-1}$Mpc}, n=0,1,2,...
\end{equation}
\begin{equation}
\pi=(m+\frac{1}{2})\mbox{$h^{-1}$Mpc}, m=0,1,2,...
\end{equation}
\begin{equation}
\mu\equiv\frac{\pi}{\sqrt{\sigma^2+\pi^2}}.
\end{equation}

Note that both the measurements and the theoretical predictions for the effective multipoles are computed using
Eq.(\ref{eq:multipole}).
We do not use the conventional definitions of multipoles to extract parameter constraints
as they use continuous integrals. 
Bias could be introduced if the definitions of multipoles are different between measurements from
data and the theoretical model.

\begin{figure}
\centering
\includegraphics[width=0.8 \columnwidth,clip,angle=-90]{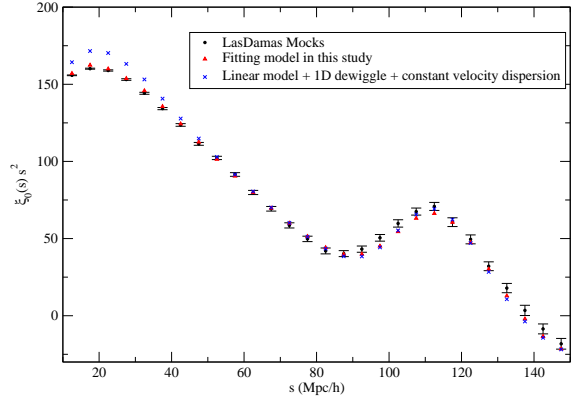}
\caption{The averaged monopole of the correlation functions of the mock catalogs (black squares) comparing to the fitting model of this study (red dots) and a simpler model (linear model + 1D dewiggle damping + constant velocity dispersion,blue crosses). The error bars are taken as $1/\sqrt{160}$ of the square roots of the diagonal elements of the covariance matrix.}
\label{fig:mono}
\end{figure}

\begin{figure}
\centering
\includegraphics[width=0.8 \columnwidth,clip,angle=-90]{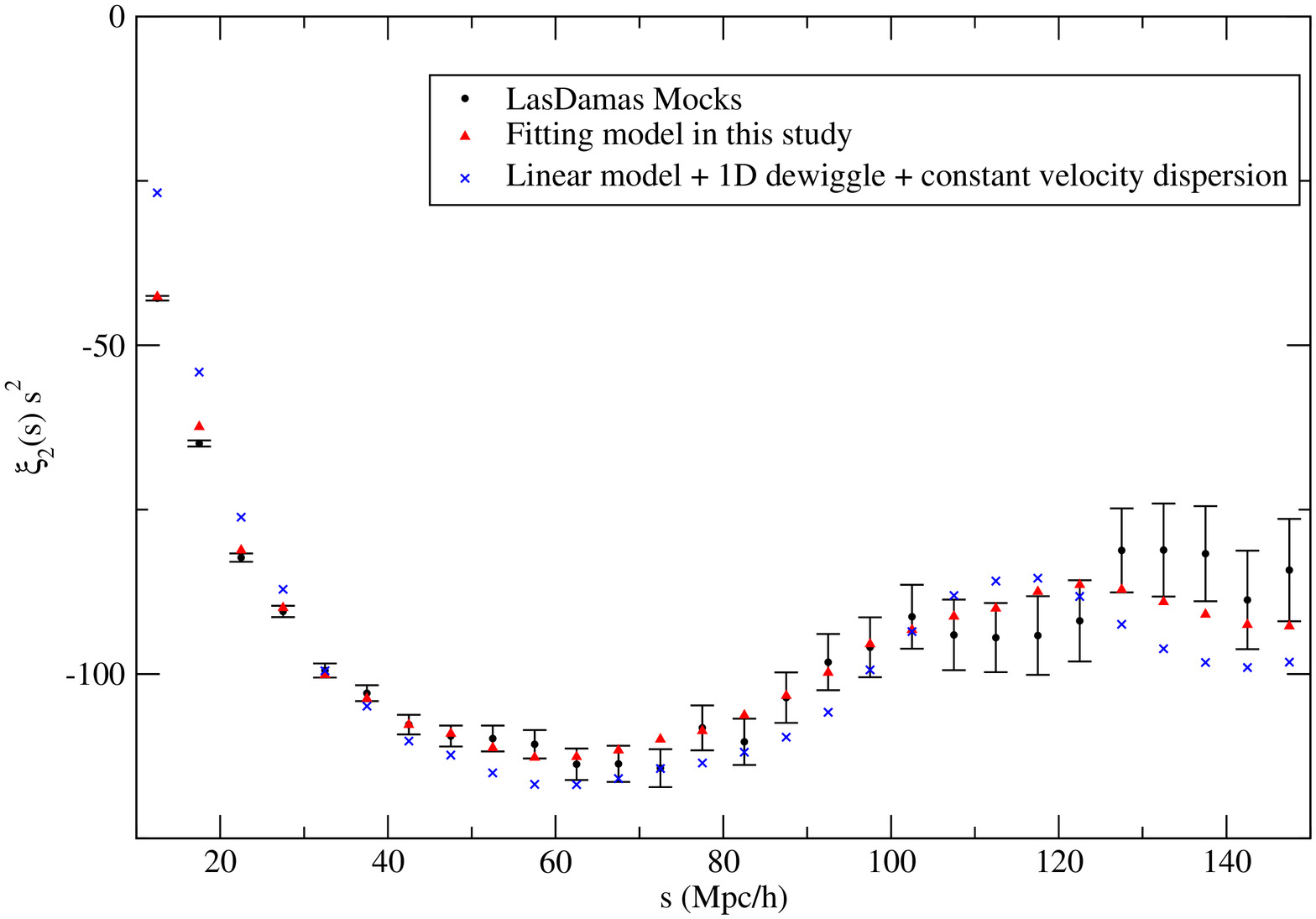}
\caption{The averaged quadrupole of the correlation functions of the mock catalogs (black squares) comparing to the fitting model of this study (red dots) and a simpler model (linear model + 1D dewiggle damping + constant velocity dispersion,blue crosses). The error bars are taken as $1/\sqrt{160}$ of the square roots of the diagonal elements of the covariance matrix.}
\label{fig:quad}
\end{figure}

\begin{figure}
\centering
\includegraphics[width=0.8 \columnwidth,clip,angle=-90]{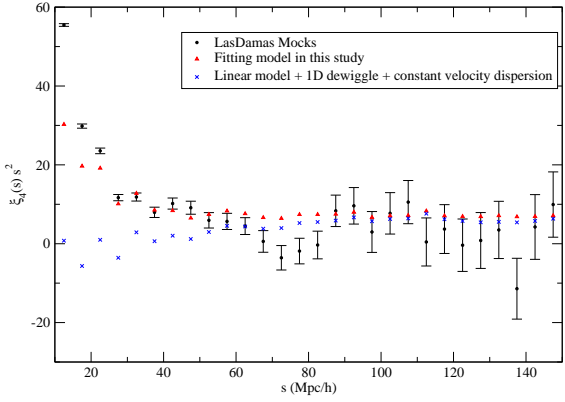}
\caption{The averaged hexadecapole of the correlation functions of the mock catalogs (black squares) comparing to the fitting model of this study (red dots) and a simpler model (linear model + 1D dewiggle damping + constant velocity dispersion,blue crosses). The error bars are taken as $1/\sqrt{160}$ of the square roots of the diagonal elements of the covariance matrix.}
\label{fig:four}
\end{figure}

Fig. \ref{fig:mono}, \ref{fig:quad}, and \ref{fig:four} show the effective 
monopole ($\hat{\xi}_0$), quadrupole ($\hat{\xi}_2$), and hexadecapole ($\hat{\xi}_4$) measured from the LasDamas mock catalogs 
comparing to our full model and a simpler model (linear model + 1D dewiggle damping + constant velocity dispersion). In Fig. \ref{fig:mono}, 
one can see how our model completely corrects the scale-dependent effects in the measured monopole. 
Fig.\ref{fig:quad} shows that our model provides a reasonable fit to the measured quadrupole.
In Fig. \ref{fig:four}, we find that angle-dependent term, 
$c_\sigma$, significantly improves the fitting of hexadecapole at small scales ($s<50h^{-1}$Mpc). However, at larger scales($s>60h^{-1}$Mpc), 
the LasDamas mocks show some oscillatory features while the theoretical models are flat. It is likely due to the dewiggle damping 
not being adequate enough to model $\xi_4$ and one might need higher order term (i.e. $\mu^4$). \
Therefore, we do not include $\xi_4$ to measure parameters in this study.

\subsection{Covariance Matrix} \label{sec:covar}

We construct the covariance matrix as
\begin{equation}
 C_{ij}=\frac{1}{N-1}\sum^N_{k=1}(\bar{X}_i-X_i^k)(\bar{X}_j-X_j^k),
\label{eq:covmat}
\end{equation}
where $N$ is the number of the mock catalogs, $\bar{X}_m$ is the
mean of the $m^{th}$ element of the vector from the mock catalog multipoles, and
$X_m^k$ is the value in the $m^{th}$ elements of the vector from the $k^{th}$ mock
catalog multipoles. The data vector ${\bf X}$ is defined by
\be
{\bf X}=\{\hat{\xi}_0^{(1)}, \hat{\xi}_0^{(2)}, ..., \hat{\xi}_0^{(N)}; 
\hat{\xi}_2^{(1)}, \hat{\xi}_2^{(2)}, ..., \hat{\xi}_2^{(N)};...\},
\label{eq:X}
\ee
where $N$ is the number of data points in each measured multipole; $N=19$ while using the scale range, $25<s<120h^{-1}$Mpc.
The length of the data vector ${\bf X}$ depends on how many multipoles are used. 

\subsection{Likelihood}
The likelihood is taken to be proportional to $\exp(-\chi^2/2)$ \citep{press92}, 
with $\chi^2$ given by
\begin{equation} \label{eq:chi2}
 \chi^2\equiv\sum_{i,j=1}^{N_{X}}\left[X_{th,i}-X_{obs,i}\right]
 C_{ij}^{-1}
 \left[X_{th,j}-X_{obs,j}\right]
\end{equation}
where $N_{X}$ is the length of the vector used, 
$X_{th}$ is the vector from the theoretical model, and $X_{obs}$ 
is the vector from the observational data (we use the mock catalogs as the observational data to test the model in this section).

As explained in \cite{Chuang:2011fy}, instead of recalculating the observed correlation function 
for different theoretical models, we rescale the theoretical correlation function to avoid rendering 
$\chi^2$ values arbitrary.
The rescaled theoretical correlation function is computed by
\begin{equation} \label{eq:inverse_theory_2d}
 T^{-1}(\xi_{th}(\sigma,\pi))=\xi_{th}
 \left(\frac{D_A(z)}{D_A^{fid}(z)}\sigma,
 \frac{H^{fid}(z)}{H(z)}\pi\right),
\end{equation}
where $\xi_{th}$ is given by eq. (\ref{eq:theory}). Hence $\chi^2$ can be rewritten as
\ba 
\label{eq:chi2_2}
\chi^2 &\equiv&\sum_{i,j=1}^{N_{X}}
 \left\{T^{-1}X_{th,i}-X^{fid}_{obs,i}\right\}
 C_{fid,ij}^{-1} \cdot \nonumber\\
 & & \cdot \left\{T^{-1}X_{th,j}-X_{obs,j}^{fid}\right\},
\ea
where $T^{-1}X_{th}$ is a vector given by eq. (\ref{eq:inverse_theory_2d}) with 
$\xi_{th}$ replaced by its effective multipoles (defined by
eq.\ (\ref{eq:multipole})), and $X^{fid}_{obs}$ is the corresponding vector from observational data measured 
assuming the fiducial model in converting redshifts to distances. See \cite{Chuang:2011fy} for a more detailed
description of our rescaling method.

\subsection{Markov Chain Monte-Carlo Likelihood Analysis}

We use CosmoMC in a Markov Chain Monte-Carlo
likelihood analysis \citep{Lewis:2002ah}. 
The parameter space that we explore spans the parameter set of
$\{H(0.35)$, $D_A(0.35)$, $\Omega_mh^2$, $\beta$, $b\sigma_8(z)$, $\Omega_bh^2$, $n_s$, $k_*$, f(0.35), $b_A$, $\sigma_{v,0}$, $c_\mu$, $c_{\sigma1}$, $c_{\sigma2}\}$. 
Only $\{H(0.35)$, $D_A(0.35)$, $\Omega_mh^2$, $\beta$, $b\sigma_8(z)\}$ are well constrained by the mock data. 

We marginalize over the other parameters, $\{\Omega_bh^2$, $n_s$,  $k_\star$, $f(0.35)$, $b_A$, $\sigma_{v,0}$, $c_\mu$, $c_{\sigma1}$, $c_{\sigma2}\}$, with flat priors over the ranges of
$\{(0.01859,0.02657)$, $(0.865,1.059)$, $(0.09,0.15)$Mpc$^{-1}h$, $(0.3,1.0)$, $(-0.2,0.2)$, $(0,500)$s$^{-1}$km, $(0,20)$, $(0,10)$, $(0.01,0.2)$Mpc$^{-2}h^2\}$, 
where the flat priors of $\Omega_b h^2$ and $n_s$ are centered on 
the measurements from WMAP7 and has width of $\pm7\sigma_{WMAP}$ (with $\sigma_{WMAP}$ from
\cite{Komatsu:2010fb}). These priors
are wide enough to ensure that CMB constraints are not double counted 
when our results are combined with CMB data \citep{Chuang:2010dv}.

\subsection{Validation of the Model Using Mock catalogs} \label{sec:validation}

We apply our method on the averaged correlation function from LasDamas SDSS LRG mock catalogs to validate our methodology. 
Table \ref{table:lasdamas} shows the measurements of 
$\{H(0.35)$, $D_A(0.35)$, $\Omega_m h^2$, , $H(0.35) \,r_s(z_d)/c$, $D_A(0.35)/r_s(z_d)$, $f(0.35)\,\sigma_8(0.35)\}$ from the averaged correlation function from
LasDamas SDSS LRG mock catalogs using $\hat{\xi}_0+\hat{\xi}_2$ and the scale range, $25<s<120h^{-1}$Mpc, comparing with the input values of the simulations.
We find the input values of the simulations are well recovered by our methodology.

\begin{table}
\begin{center}
\begin{tabular}{crr}\hline
&$25<s<120$ &input value \\ \hline
$	H(0.35)	$&\ \ $82.8\pm11$ &\ \  $81.79$ \\
$	D_A(0.35)$&\ \  $1023 \pm 77$ &\ \  $1032.8$ \\
$	\Omega_m h^2	$&\ \  $0.120\pm0.020$ &\ \  $0.1225$ \\
$	H(0.35) \,r_s(z_d)/c	$&\ \  $0.0444 \pm 0.0054$ &\ \  $0.0434$ \\
$	D_A(0.35)/r_s(z_d)	$&\ \  $6.35 \pm 0.45$ &\ \  $6.48$ \\
$	f(0.35)*\sigma_8(0.35)	$&\ \  $0.445\pm0.097$ &\ \  $0.437$ \\
\hline
\end{tabular}
\end{center}
\caption{
The mean and standard deviation of 
$\{H(0.35)$, $D_A(0.35)$, $\Omega_m h^2$,
$H(0.35) \,r_s(z_d)/c$, $D_A(0.35)/r_s(z_d)$, $f(0.35)\,\sigma_8(0.35)\}$ from the averaged correlation function from
LasDamas SDSS LRG mock catalogs using $\hat{\xi}_0+\hat{\xi}_2$ and the scale range, $25<s<120h^{-1}$Mpc, comparing with the input values of the simulations.
One can see that the input values of the simulations are well recovered by our methodology.
The unit of $H$ is $\Hunit$. The unit of $D_A$ and $r_s(z_d)$ is $\rm Mpc$.
} \label{table:lasdamas}
\end{table}

\section{Measurements from SDSS DR7 LRG} 
\label{sec:results_sdss}

Table \ref{table:sdss} lists the mean and rms variance of the parameters, $\{H(0.35)$, $D_A(0.35)$, $\Omega_m h^2$, $\beta$, $	b\,\sigma_8(z)	$
$H(0.35) \,r_s(z_d)/c$, $D_A(0.35)/r_s(z_d)$, $f(z)\,\sigma_8(z)\}$,
derived in an MCMC likelihood analysis from the measured $\hat{\xi}_0+\hat{\xi}_2$ of the correlation function
of the SDSS LRG sample with the scale range, $25<s<120\,h^{-1}$Mpc and $40<s<120\,h^{-1}$Mpc.
Table \ref{table:covar_matrix_25} and \ref{table:covar_matrix_40} gives the corresponding normalized covariance matrices.

While we are modeling the correlation function on small scales,
the uncertainties would still become larger when smaller scales are included.
Although one could obtain tighter constraints by using very small 
scales, to be conservative, we only fit the measurements using scales 
larger than 25 $h^{-1}$Mpc and check the consistency with the measurements 
using the scales larger than 40 $h^{-1}$Mpc.

Our measurements are consistent between two scale ranges considered which shows no hint of systematics. 
We choose the results using the scale range, $25<s<120\,h^{-1}$Mpc, as our fiducial results.
As expected, the constraints become tighter when including smaller scales. 
Notice that the correlations between $\Omega_m h^2$ and $\{H(0.35)r_s(z_d)/c$, $D_A(0.35)/r_s(z_d)\}$
also increase. It is due to the fact that our measurements gain more constraining power from the overall 
shape beyond the BAO peak region.

\begin{table}
\begin{center}
\begin{tabular}{crr}\hline
&$25<s<120$ &$40<s<120$ \\ \hline
$	H(0.35)	$&\ \ $82.7\pm8.4$ &\ \  $79\pm12$ \\
$	D_A(0.35)$&\ \  $1036 \pm 79$ &\ \  $1039\pm113$ \\
$	\Omega_m h^2	$&\ \  $0.1226\pm0.025$ &\ \  $0.101\pm0.017$ \\
$	\beta	$&\ \  $0.388\pm0.081$ &\ \  $0.426\pm0.15$ \\
$	b\,\sigma_8(z)	$&\ \  $1.110\pm0.079$ &\ \  $1.038\pm0.095$ \\
\hline
$	H(0.35) \,r_s(z_d)/c	$&\ \  $0.0433 \pm 0.0042$ &\ \  $0.0432\pm0.0064$ \\
$	D_A(0.35)/r_s(z_d)	$&\ \  $6.59 \pm 0.46$ &\ \  $6.30\pm0.65$ \\
$	f(0.35)\,\sigma_8(0.35)	$&\ \  $0.429\pm0.089$ &\ \  $0.438	\pm0.14$ \\
\hline
$\chi^2$/d.o.f.&\ \ $1.07-1.46$&\ \ $1.05-1.57$\\
\hline
\end{tabular}
\end{center}
\caption{
The mean and standard deviation of 
$\{H(0.35)$, $D_A(0.35)$, $\Omega_m h^2$, $\beta$, $b\,\sigma_8(z)$, 
$H(0.35) \,r_s(z_d)/c$, $D_A(0.35)/r_s(z_d)$, $f(0.35)\,\sigma_8(0.35)\}$ from SDSS DR7 LRGs using $\hat{\xi}_0+\hat{\xi}_2$ and the scale ranges, $25<s<120h^{-1}$Mpc and $40<s<120h^{-1}$Mpc.
We report the minimum and maximum $\chi^2$ per degree of freedom since there are many fitting parameters are not well constrained.
The unit of $H$ is $\Hunit$. The unit of $D_A$ and $r_s(z_d)$ is $\rm Mpc$.
} \label{table:sdss}
\end{table}

\begin{table*}
\begin{center} 
\begin{tabular}{crrrrrrrrrr}\hline
       &$H(0.35)$ &$D_A(0.35)$   &$\Omega_mh^2$ &$\beta$ &$b\sigma_8(0.35)$ &$H(0.35) \,r_s(z_d)/c$ &$D_A(0.35)/r_s(z_d)$ &$f(0.35)\sigma_8(0.35)$ \\ \hline
$H(0.35)$&\ \ 	1.0000	&\ \	-0.0069	&\ \	0.3361	&\ \	0.4316	&\ \	-0.0867	&\ \	0.8478	&\ \	0.2602	&\ \	0.3995	\\
$D_A(0.35)$&\ \ 	-0.0069	&\ \	1.0000	&\ \	-0.4422	&\ \	0.2151	&\ \	-0.0338	&\ \	0.2539	&\ \	0.7257	&\ \	0.2029	\\
$\Omega_mh^2$&\ \	0.3361	&\ \	-0.4422	&\ \	1.0000	&\ \	0.1189	&\ \	0.6116	&\ \	-0.1937	&\ \	0.2664	&\ \	0.3278	\\
$\beta$&\ \ 	0.4316	&\ \	0.2151	&\ \	0.1189	&\ \	1.0000	&\ \	-0.1693	&\ \	0.3983	&\ \	0.3011	&\ \	0.9400	\\
$b\sigma_8(0.35)$ &\ \	-0.0867	&\ \	-0.0338	&\ \	0.6116	&\ \	-0.1693	&\ \	1.0000	&\ \	-0.3864	&\ \	0.3746	&\ \	0.1705	\\
$H(0.35) \,r_s(z_d)/c$&\ \	0.8478	&\ \	0.2539	&\ \	-0.1937	&\ \	0.3983	&\ \	-0.3864	&\ \	1.0000	&\ \	0.1195	&\ \	0.2626	\\
$D_A(0.35)/r_s(z_d)$&\ \	0.2602	&\ \	0.7257	&\ \	0.2664	&\ \	0.3011	&\ \	0.3746	&\ \	0.1195	&\ \	1.0000	&\ \	0.4299	\\
$f(0.35)\sigma_8(0.35)$&\ \	0.3995	&\ \	0.2029	&\ \	0.3278	&\ \	0.9400	&\ \	0.1705	&\ \	0.2626	&\ \	0.4299	&\ \	1.0000	\\
\hline
\end{tabular}
\end{center}
\caption{Normalized covariance matrix of the measured and derived parameters, $\{H(0.35)$, $D_A(0.35)$, $\Omega_m h^2$, $\beta,$, $b\sigma_8(0.35)$,
$H(0.35) \,r_s(z_d)/c$, $D_A(0.35)/r_s(z_d)$, $f(0.35)\sigma_8(0.35)\}$ from SDSS DR7 LRGs using $\hat{\xi}_0+\hat{\xi}_2$ and the scale range, $25<s<120h^{-1}$Mpc.}
 \label{table:covar_matrix_25}
\end{table*}

\begin{table*}
\begin{center} 
\begin{tabular}{crrrrrrrrrr}\hline
       &$H(0.35)$ &$D_A(0.35)$   &$\Omega_mh^2$ &$\beta$ &$b\sigma_8(0.35)$ &$H(0.35) \,r_s(z_d)/c$ &$D_A(0.35)/r_s(z_d)$ &$f(0.35)\sigma_8(0.35)$ \\ \hline
$H(0.35)$&\ \ 	1.0000	&\ \	0.1050	&\ \	0.2602	&\ \	0.4394	&\ \	-0.1514	&\ \	0.9566	&\ \	0.2280	&\ \	0.4217	\\
$D_A(0.35)$&\ \ 	0.1050	&\ \	1.0000	&\ \	-0.2925	&\ \	0.3573	&\ \	0.1515	&\ \	0.2001	&\ \	0.9164	&\ \	0.4097	\\
$\Omega_mh^2$&\ \	0.2602	&\ \	-0.2925	&\ \	1.0000	&\ \	-0.0115	&\ \	0.5157	&\ \	-0.0164	&\ \	0.0903	&\ \	0.1284	\\
$\beta$&\ \ 	0.4394	&\ \	0.3573	&\ \	-0.0115	&\ \	1.0000	&\ \	-0.2872	&\ \	0.4593	&\ \	0.3711	&\ \	0.9627	\\
$b\sigma_8(0.35)$&\ \	-0.1514	&\ \	0.1515	&\ \	0.5157	&\ \	-0.2872	&\ \	1.0000	&\ \	-0.2911	&\ \	0.3396	&\ \	-0.0357	\\
$H(0.35) \,r_s(z_d)/c$&\ \	0.9566	&\ \	0.2001	&\ \	-0.0164	&\ \	0.4593	&\ \	-0.2911	&\ \	1.0000	&\ \	0.2068	&\ \	0.4037	\\
$D_A(0.35)/r_s(z_d)$&\ \	0.2280	&\ \	0.9164	&\ \	0.0903	&\ \	0.3711	&\ \	0.3396	&\ \	0.2068	&\ \	1.0000	&\ \	0.4758	\\
$f(0.35)\sigma_8(0.35)$&\ \	0.4217	&\ \	0.4097	&\ \	0.1284	&\ \	0.9627	&\ \	-0.0357	&\ \	0.4037	&\ \	0.4758	&\ \	1.0000	\\
\hline
\end{tabular}
\end{center}
\caption{Normalized covariance matrix of the measured and derived parameters, $\{H(0.35)$, $D_A(0.35)$, $\Omega_m h^2$, $\beta,$, $b\sigma_8(0.35)$,
$H(0.35) \,r_s(z_d)/c$, $D_A(0.35)/r_s(z_d)$, $f(0.35)\sigma_8(0.35)\}$ from SDSS DR7 LRGs using $\hat{\xi}_0+\hat{\xi}_2$ and the scale range, $40<s<120h^{-1}$Mpc.}
 \label{table:covar_matrix_40}
\end{table*}

\section{Conclusion and Discussion} 

We have presented and validated a simple and efficient phenomenological model for the two-dimensional two-point 
galaxy correlation function that works well over a wide range of scales, from large scales down to 
small scales not used in our previous work (where we restricted ourselves to scales larger than
40$\,h^{-1}$Mpc). Applying this model to the SDSS LRGs over the scale range of $25<s<120h^{-1}$Mpc, 
We obtain the measurements: $H(z)r_s(z_d)/c=0.0433\pm 0.0042$, $D_A(z)/r_s(z_d)=6.59\pm 0.46$, and 
$f(z)\sigma_8(z)=0.429\pm 0.089$ at $z=0.35$, which summarize the cosmological constraints extracted from the SDSS DR7 LRG sample.
We also provide the covariance matrix needed to use these measurements (see Table 4).

Our model incorporates the overall nonlinear effects via the use of the ``dewiggled'' galaxy power spectrum,
as in \cite{Chuang:2011fy}, but we now include the enhanced damping along the line of sight 
(see Eqs.(\ref{eq:pk_lin})-(\ref{eq:pk_dw})). We also introduce a much efficient way to compute this model which is crucial for MCMC analysis.
On small scales, the nonlinear effect and scale-dependent galaxy bias 
are degenerate, and we model these as an overall scale-dependent correction.
Most significantly, we allow the RSD to be scale and direction dependent in our model. 
Our model provides excellent fit to mock data (see Fig.\ref{fig:twod}).

We expect our methodology and results to be useful in tightening dark energy and gravity constraints
from the full analysis of current and future galaxy clustering data. 

\label{sec:conclusion}

\section*{Acknowledgements}
We are grateful to the LasDamas project              
for making their mock catalogs publicly available.
The computing for this project was performed at the OU 
Supercomputing Center for Education and Research (OSCER) at the University of 
Oklahoma (OU).
This work used the Extreme Science and Engineering Discovery Environment (XSEDE), which is supported by National Science Foundation grant number OCI-1053575. 
This work was supported by DOE grant DE-FG02-04ER41305.
C.C. was also supported by the Spanish MICINN’s Consolider-Ingenio 2010 Programme under grant MultiDark CSD2009-00064 and grant AYA2010-21231,
the Comunidad de Madrid under grant HEPHACOS S2009/ESP-1473,
and the Spanish MINECO’s “Centro de Excelencia Severo Ochoa” Programme under grant SEV-2012-0249.

Funding for the Sloan Digital Sky Survey (SDSS) has been provided by the Alfred P. Sloan Foundation, the Participating Institutions, the National Aeronautics and Space Administration, the National Science Foundation, the U.S. Department of Energy, the Japanese Monbukagakusho, and the Max Planck Society. The SDSS Web site is http://www.sdss.org/.

The SDSS is managed by the Astrophysical Research Consortium (ARC) for the Participating Institutions. The Participating Institutions are The University of Chicago, Fermilab, the Institute for Advanced Study, the Japan Participation Group, The Johns Hopkins University, Los Alamos National Laboratory, the Max-Planck-Institute for Astronomy (MPIA), the Max-Planck-Institute for Astrophysics (MPA), New Mexico State University, University of Pittsburgh, Princeton University, the United States Naval Observatory, and the University of Washington. 
\setlength{\bibhang}{2.0em}

\label{lastpage}

\end{document}